\documentstyle[twocolumn,epsf,amssymb,amsmath]{mn}
\newcommand{\lcdm}{$\Lambda$CDM }
\newcommand{\mincir}{\raise
-3.truept\hbox{\rlap{\hbox{$\sim$}}\raise4.truept\hbox{$<$}\ }}
\newcommand{\magcisr}{\raise
-3.truept\hbox{\rlap{\hbox{$\sim$}}\raise4.truept\hbox{$>$}\ }}
\newcommand{\minmag}{\raise
-3.truept\hbox{\rlap{\hbox{$<$}}\raise5.truept\hbox{$<$}\ }}
\newcommand{\be}{\begin{equation}}
\newcommand{\ee}{\end{equation}}

\newcommand{\ba}{\begin{eqnarray}}
\newcommand{\ea}{\end{eqnarray}}
\newcommand{\brr}{\begin{array}}
\newcommand{\err}{\end{array}}
\newcommand{\bc}{\begin{center}}
\newcommand{\ec}{\end{center}}

\renewcommand{\baselinestretch}{1.0}

\title[Testing dynamical vacuum models with Planck spectrum]
{Testing dynamical vacuum models with CMB power spectrum from Planck}
\author[Pavlina Tsiapi, Spyros Basilakos] {P. Tsiapi$^{1}$, S. Basilakos$^{2,3}$\\
\vspace{0.1cm} $^{1}$ National Technical University of Athens, School of Applied Mathematical and Physical Sciences, Iroon Polytechneiou 9,\\
 15780, Athens, Greece \\
$^{2}$ Academy of Athens, Research Center for Astronomy \& Applied
Mathematics, Soranou Efessiou 4, 11-527, Athens, Greece \\
$^{3}$ National Observatory of Athens, Lofos Nymfon, 11851 Athens, Greece
  }

\begin{document}
  
\maketitle

\begin{abstract}
The cosmic expansion  
is computed for various dynamical vacuum models $\Lambda(H)$  
and confronted 
to the  
Cosmic Microwave Background (CMB) power spectrum from 
Planck. We also combined CMB in a joint analysis with other 
probes to place constraints on the cosmological parameters of the dynamical
vacuum models. We find that all $\Lambda(H)$ 
models are very efficient and in very good agreement with the data. 
Considering that the interaction term of the dark sector 
is given in terms of matter and radiation densities  
we find that the corresponding $\Lambda(H)$ model
shows a small but non-zero deviation from $\Lambda$ 
cosmology, nevertheless the confidence level is close
to $\sim 2.5\sigma$.


\end{abstract}

\vspace{1.0cm}

\section{Introduction}

The comprehensive analysis of a large family of observational data 
indicate that around $\sim 95\%$ of the Universe content
corresponds to unknown sectors, usually called 
dark matter (around $\sim 25\%$) and 
dark energy (around $\sim 70\%$). Dynamically, the latter component 
plays a key role in cosmic expansion because it 
is responsible for the outstanding 
phenomenon of cosmic acceleration 
(Perlmutter et al. 1998; Riess et al. 1998; Astier et al. 2007; Komatsu et al. 2011; Suzuki et al. 2012; Ade et al. 2016; Aghanim et al. 2018).
It is worth mentioning that the nature of dark energy (hereafter DE) 
has been a topic 
of interest ever since its first appearance in 
Einstein's equations as a cosmological constant.
Even though the term was briefly discarded as 
unnecessary, the cosmological constant is to date the most 
favored candidate as a dark energy which 
coexists with cold dark matter (CDM) and ordinary baryonic matter
(see Peebles \& Ratra 2003, for review). These components form the so called 
$\Lambda$CDM model which fits accurately the current 
cosmological data.

Although the $\Lambda$CDM model is 
considered as a successful cosmic scenario, it is not without its problems (Weinberg 1989; Padmanabhan 2003; Perivolaropoulos 2008; Padilla 2015). 
Examples of theoretical impairments of the model include 
the fine tuning and coincidence problems.
The fine tuning problem reflects the gap between 
the expected (Planck natural unit) vacuum energy density 
$\rho_{\rm vac}$, which, using 
quantum field theory (QFT), is calculated at a remarkable $\sim 120$ 
orders of magnitude larger that the observed value of $\rho_{\Lambda}$
at the present time.
On the other hand, the coincidence problem sources 
from the approximate equality of 
$\rho_{m}$  and $\rho_{\Lambda}$ prior to the present time, even though the former 
is a dynamical quantity and the latter is a constant. 


These problems have given rise to a large body of 
cosmological models which mainly extends the traditional 
Einstein-Hilbert action of general relativity using either 
a new field 
(Horndeski 1974; Brans \& Dicke 1961; Nicolis, Rattazzi \& Trincherini 2009; Urena - Lopez 2016), or a modified gravity theory that
increases the number of degrees of freedom
(Buchdahl 1970; Bengochea \&  Ferraro 2009; Sotiriou \& Faraoni 2010; Clifton, Ferreira, Padilla \& Skordis 2012; Nunes et. al. 2016;  Nunes, Pan \& Saridakis 2016, see also Copeland, Sami \& Tsujikawa 2006; Caldwell \& Kamionkowski 2009; Amendola \& Tsujikawa 2010).
Among the class of DE models, 
the introduction of a dynamical vacuum, $\Lambda(t)$, is perhaps the
simplest modification of the Einstein-Hilbert action 
towards alleviating the aforementioned theoretical issues
(see Basilakos 2009; Basilakos, Lima \& Sola 2013; Perico, Lima, Basilakos \& Sola 2013; Basilakos, Mavromatos \& Sola 2016). 
Here, the time dependence is not introduced via the equation 
of state (EoS) parameter $w_\Lambda$ which, like the $\Lambda$CDM, is strictly 
set to $w_\Lambda = -1$, but is inherited to the 
pressure through $p_{\Lambda}(t) = -\rho_{\Lambda}(t)$.
Notice, that the idea to deal with 
a vacuum which varies with cosmic time (or redshift) 
has a long history in cosmology and it is 
perfectly allowed by the cosmological principle
(Ozer \& Taha 1986; Bertolami 1986; Chen \& Wu 1990; Carvalho, Lima \& Waga 1992; Lima \& Maia 1993; Salim \& Waga 1993; Arcuri \& Waga 1994; Lima \& Carvalho 1994; Lima \& Trodden 1996; John \& Joseph 2000; Lima, Maia \& Pires 2000; Novello, Barcelos-Neto \& Salim 2001; Vishwakarma 2001; Cunha, Lima \& Pires 2002; Schutzhold 2002a, 2002b; Aldrovandi, Beltran Almeida \& Pereira 2005; Carneiro \& Lima 2005; Pan 2018).

Usually, the dynamical vacuum energy density 
$\rho_\Lambda$ evolves slowly as a power series
of the Hubble rate (for a review 
see Shapiro \& Sola 2002; Grande et. al. 2011; Sola 2013, 2014; Gomez, Sola \& Basilakos 2015). 
In this scenario, the decaying vacuum energy density has 
an interesting feature, namely it predicts that the spacetime 
emerges from a non-singular initial de Sitter vacuum stage, hence 
the phase of the universe changes in a smooth way  
from inflation to a radiation epoch ("graceful exit").
Then the universe enters into the  
dark-matter and vacuum-dominated phases, before
finally, entering to a late-time de Sitter phase (Basilakos et. al. 2013, 2016; Perico et. al. 2013).
From the observational view point, recently 
(Sola, Gomez - Valent \& de Cruz Perez 2017a, 2017b, 2018) tested the performance of the running
vacuum models against the latest cosmological data and they found that the
$\Lambda(H)$ models are favored over the usual $\Lambda$CDM model at
$\sim 3-4\sigma$ statistical level (see also Zhao et al. 2017).
These results have led to growing interest in $\Lambda(H)$ 
cosmological models.

In this work we attempt to study the performance
of 
various dynamical 
vacuum models at the
expansion level. 
Specifically, a likelihood 
analysis, involving the Planck CMB power spectrum (Ade et. al. 2016), is 
implemented in order to constrain the $\Lambda(H)$ models. 
Up to now, for these models only the set of the CMB shift 
parameters have been used in order to test 
the dynamical vacuum scenarios close to recombination. 
Then we 
combine CMB in an overall likelihood analysis with other cosmological probes
(SNe type Ia, BAOs, $H_{0}$) in order 
to extract the probability distribution (via Monte Carlo MCMC method) 
of the solutions for a large set of cosmological parameters, 
as well as, to search for deviations from the concordance 
$\Lambda$CDM model for which $\Lambda(H)= \text{const}$.

The structure of our paper is as follows: in section 2 we provide  
a brief introduction of the running vacuum cosmology and we present 
the most popular $\Lambda(H)$ models 
that have appeared in the literature. 
In section 3
we discuss the methodology and the cosmological data that we
utilize and we perform a
detailed statistical analysis aiming to provide the 
corresponding best fit values and contour plots for the current 
$\Lambda(H)$ models.
Finally, we discuss our results in the conclusions.

\section{Background expansion in running vacuum models}
\label{sec:RVcosmology}
In this section we briefly describe the main features of the dynamical 
vacuum models for which $\Lambda$ is not constant but evolves 
with cosmic time. 
This is perfectly allowed by the cosmological principle
embedded in the FLRW metric (Perico et. al. 2013). In general, if we model
the expanding universe as a mixture of perfect fluids
$N=1,2,..$ then the total energy momentum tensor is given by
\begin{equation}\label{Tmunu}
{T}^\mu_\nu = \sum_N {T}^{\mu, N}_\nu=\sum_N\left[ -
p_N\,\delta^\mu_\nu+\big(\rho_N + p_N\,\big)\,U^{\mu, N}\,U_{\nu}^N\right]\,,
\end{equation}
where $U_{\mu}^{N}$ is the 4-velocity field. 
In this case the components of $T^{\mu}_{\nu}$ are written as
\begin{equation}\label{Tcomp}
{T}^0_{0}=\sum_N \rho_N \equiv \rho_{\rm T} \,,\ \
{T}^i_j=-\sum_N p_N \,\delta^i_j\equiv -p_{\rm T}\,\delta^i_j \,,
\end{equation}
where the quantities $p_{\rm T}$  and $ \rho_{\rm T}$ 
are the total pressure and energy density in the 
comoving frame $(U_N^0,U_N^i)=(1,0)$, respectively.
The next step is to apply the covariant local conservation law for the mixture,
namely $\nabla_\mu {T}^{\mu\nu} = 0$. 
Inserting Eq.(\ref{Tmunu}) into the latter expression and with the aid 
of the relation $U_{\nu}^N\nabla_{\mu}U^{\nu}_N=0$ (which
comes from the fact that for any four-velocity vector, we
have $U^{\mu}_N\,U_{\mu}^N=1$) we arrive at (Grande, Pelinson \& Sola 2009):

\begin{equation}\label{conserv3}
\sum_N\left[U_N^\mu\,\nabla_{\mu}\,\rho_N+(\rho_N+p_N)\nabla_\mu
{U}^\mu_N\right]=0\,.
\end{equation}
For a Friedmann-Lema\^{\i}tre-Robertson-Walker (FLRW) metric, it is
easy to check that for a comoving frame (${U}^\mu_N =
\delta^{\mu}_{0}$), one obtains:
\begin{equation}\label{nablaU3H}
\nabla_\mu {U}^\mu_N=3\,H\,\ \ \ (N=1,2,...)\,,
\end{equation}
and thus Eq.(\ref{conserv3}) reduces to
\begin{equation}\label{EnergyCons}
\sum_N\left[\, \dot{\rho}_N + 3 H (\rho_N+p_N)\,\right]
=0\,,
\end{equation}
where $H={\dot a}/a$ is the Hubble parameter and $a(t)$ 
is the scale factor of the universe normalized to unity 
at the present epoch. For the rest of the paper we 
focus on the spatially flat FLRW metric.

In the aforementioned discussion we did not address  
the physics of the fluids involved. 
The total density $\rho_{T}$ receives contributions 
from non-relativistic matter (cold dark matter and baryons) 
$\rho_{m}=\rho_{b}+\rho_{\rm dm}$ ($p_{m}=0$), radiation $\rho_{r}$ 
($p_{r}=\rho_{r}/3$) and vacuum 
$\rho_{\Lambda}$ ($p_{\Lambda}=-\rho_{\Lambda}$). 
In this context the index $N$ specifies the 
specific components of the cosmic fluid, namely 
$\{{\rm dm},b,r,\Lambda \}$.
Assuming that baryons and radiation 
are self-conserved, namely the corresponding densities evolve in
the nominal way $\rho_{b}=\rho_{b0}a^{-3}$ and $\rho_{r}=\rho_{r0}a^{-4}$ 
the overall conservation law (\ref{EnergyCons}) becomes 
\begin{equation}
\label{lambdavar}
{\dot\rho}_{\Lambda}+{\dot\rho}_{\rm dm}+3H\rho_{dm}=0
\end{equation}
or 
\begin{equation}
\label{eq:Q1}
{\dot \rho}_{\rm dm} + 3H\rho_{\rm dm} = Q
\end{equation}
\begin{equation}
\label{eq:Q2}
{\dot \rho}_{\Lambda} =-Q.
\end{equation}
Notice that $Q$ is the interaction term between dark matter and running vacuum, 
such that a small amount of vacuum decays into dark matter or vice versa. 
It is worth mentioning that the above expression 
is the outcome of imposing the covariant conservation of the
total energy density of the combined system of matter and vacuum, hence 
is a direct consequence of the Bianchi identity in the context of 
general relativity. 
Of course in the case 
of the concordance $\Lambda$CDM model ($\rho_{\Lambda}$=const., $Q=0$), 
we recover the standard dark matter
conservation law ${\dot\rho}_{\rm dm}+3H\rho_{\rm dm}=0$.

Within this framework, the Friedmann equations of the system
formed by the above fluid components are given by 
(see Sola et. al. 2017a, and references therein):
\begin{equation}
\label{eq:F1}
H^2 = \frac{8\pi G}{3} (\rho_m+\rho_r+\rho_\Lambda)
\end{equation}
\begin{equation}
\label{eq:F2}
2{\dot H} +3H^2 = -8\pi G\left(\frac{1}{3}\rho_r-\rho_\Lambda\right).
\end{equation}

\subsection{Specific $\Lambda(H)$ models}
Now let us briefly present the running vacuum models studied in this
article. For each
one of these models we provide the term of interaction $Q$ 
and thus we calculate the evolution of main cosmological quantities, namely
$\rho_{\Lambda}(a)$, $\rho_{\rm dm}(a)$ and $H(a)$. Notice that the 
baryon and the radiation densities obey the standard 
laws, namely $\rho_{b}(a) \propto a^{-3}$ and $\rho_{r}(a) \propto a^{-4}$ 
respectively. 
Specifically, the current $\Lambda(H)$ models read as follows.


The first model under consideration is the 
running vacuum model as described in (Basilakos 2009; Basilakos et. al. 2013, 2016; Perico et. al. 2013) 
(hereafter $\Lambda(H)$CDM$_{1}$). 
In this case we have $Q=\nu H (3\rho_{m}+4\rho_{r})$, hence 
solving the system of equations (\ref{eq:Q1})-(\ref{eq:Q2})
one finds an expression for the evolution of both densities:

\begin{equation}
\begin{split}
\rho_{\rm dm} = \rho_{dm,0}a^{-3(1-\nu)}+\rho_{b,0}(a^{3(1-\nu)}-a^{-3})\\+\frac{4\nu}{1+3\nu}\rho_{r,0}\left(a^{3(1-\nu)}-a^{-4}\right)
\end{split}
\end{equation}
\begin{equation}
\begin{split}
\rho_\Lambda = \rho_{\Lambda,0}+\frac{\nu \rho_{m,0}}{1-\nu}(a^{3(1-\nu)}-1)\\+\frac{\nu \rho_{r,0}}{1-\nu}\left( \frac{1-\nu}{1+3\nu}a^{-4}+\frac{4\nu}{1+3\nu}a^{-3(1-\nu)}-1\right),
\end{split}
\end{equation}
while the normalized Hubble parameter $E(a)=H(a)/H_{0}$ is given by
\begin{equation}
\begin{split}
\label{eq:rvmh2}
E^2(a) = 1+\frac{\Omega_m}{1-\nu}\left( a^{-3(1-\nu)}-1 \right)\\
+\frac{\Omega_r}{1-\nu}\left( \frac{1-\nu}{1+3\nu}a^{-4}+\frac{4\nu}{1+3\nu}a^{-3(1-\nu)}-1 \right).
\end{split}
\end{equation} 

The second phenomenological 
model that we take into account is that with $Q=3\nu H \rho_{\rm dm}$ 
(hereafter $\Lambda(H)$CDM$_{2}$).
In this context, we have 
\be
\label{eq:dmrdm}
\rho_{dm}= \rho_{dm,0}a^{-3(1-\nu)}
\ee
\be
\label{eq:dmrl}
\rho_\Lambda = \rho_{\Lambda,0}+\frac{\nu\rho_{dm,0}}{1-\nu}\left( a^{-3(1-\nu)}-1\right)
\ee
and 
\be
\label{eq:dmh2}
E^2= 1+\Omega_b(a^{-3}-1)+\frac{\Omega_{dm}}{1-\nu}\left( a^{-3(1-\nu)}-1\right)+\Omega_r(a^{-4}-1).
\ee

The final vacuum model consists of
$Q=3\nu H \rho_{\Lambda}$ (hereafter $\Lambda(H)$CDM$_{3}$). Within this 
framework the basic cosmological quantities become

\be
\label{eq:lrdm}
\rho_{dm} =  \rho_{dm,0}+\frac{\nu\rho_{dm,0}}{1-\nu}\left( a^{-3\nu}-a^{-3}\right)
\ee

\be
\label{eq:lrl}
\rho_{\Lambda} = \rho_{\Lambda,0}a^{-3\nu}
\ee
\be
\begin{split}
	\label{eq:lh2}
	E^2= \frac{a^{-3\nu}-\nu a^{-3}}{1-\nu}+\frac{\Omega_m}{1-\nu}(a^{-3}-a^{-3\nu})\\
	+\Omega_r\left(a^{-4}+\frac{\nu}{1-\nu}a^{-3}-\frac{a^{-3\nu}}{1-\nu}\right).
\end{split}
\ee


It is important to note that for $\nu=0$
the aforementioned equations boil down to those of $\Lambda$CDM, as 
they should.

\section{Fitting running vacuum models to the Planck CMB spectrum}

For this analysis, the publicly available CAMB code was 
modified to admit dynamical vacuum models, and used in 
combination with the MCMC package to restrain 
the usual set of cosmological parameters of the standard model 
for Cosmology, with the addition of the dynamical vacuum parameter $\nu$. 
We use the Planck satellite, 2015 release which includes 
the CMB power spectrum, TT, TE, EE + lowP (Ade et. al., 2016). 
For completeness we also give the constrains of 
the $\Lambda$CDM model. 
Notice that the parameter space is 
$\{ \Omega_m, \sigma_{8},h,\nu, n_s,\tau\}$, where 
$\sigma_{8}$ is the mass variance at $8h^{-1}$Mpc, $h=H_{0}/100$, 
$n_{s}$ spectral index, $\tau$ is the optical depth. 
Notice that in CAMB, the density of photons is calculated through $\rho_r=\frac{4\sigma_BT_{\rm cmb}^4}{c^3}$, with $\sigma_B$ being 
the Stefan - Boltzmann constant, and $T_{\rm cmb}$ the temperature of the CMB.

\begin{figure}
	\mbox{\epsfxsize=8.5cm \epsffile{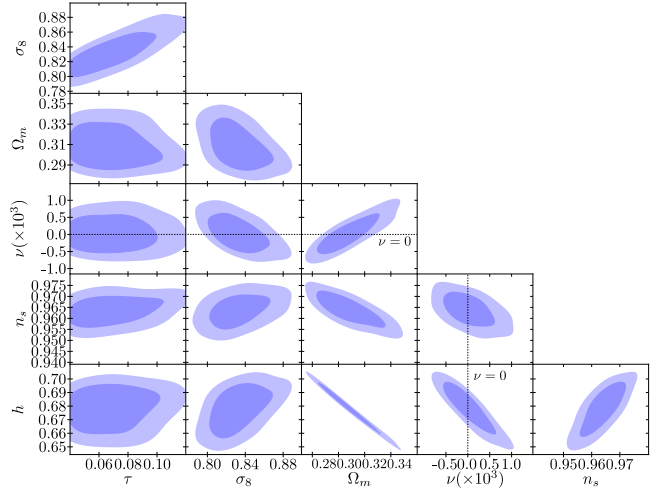}} \caption{2D likelihood contours for the $\Lambda(H)$CDM$_{1}$ vacuum model. 
	We present $1\sigma$ and $2\sigma$ likelihood contours of all the sampled parameters when testing against the CMB full spectrum. 
	The straight line corresponds to $\nu=0$.}
	\label{fig:Q1}
\end{figure}

\begin{figure}
	\mbox{\epsfxsize=8.5cm \epsffile{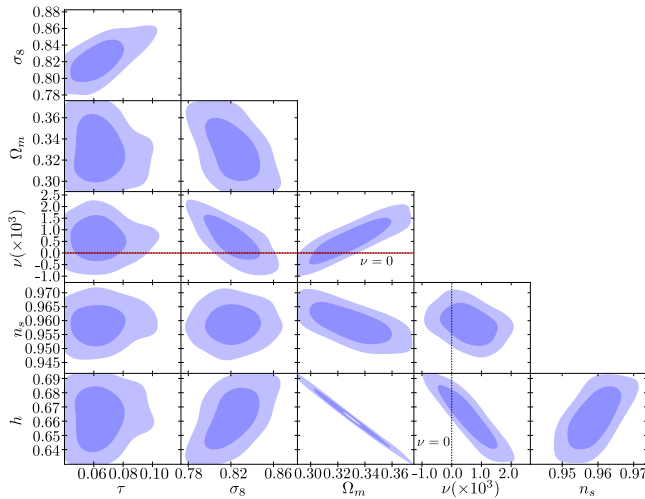}} \caption{The CMB spectrum analysis, repeated for the $\Lambda(H)$CDM$_2$ model. 
	Again, the straight line corresponds to $\nu=0$.}
	\label{fig:Q2}
\end{figure}

\begin{figure}
	\mbox{\epsfxsize=8.5cm \epsffile{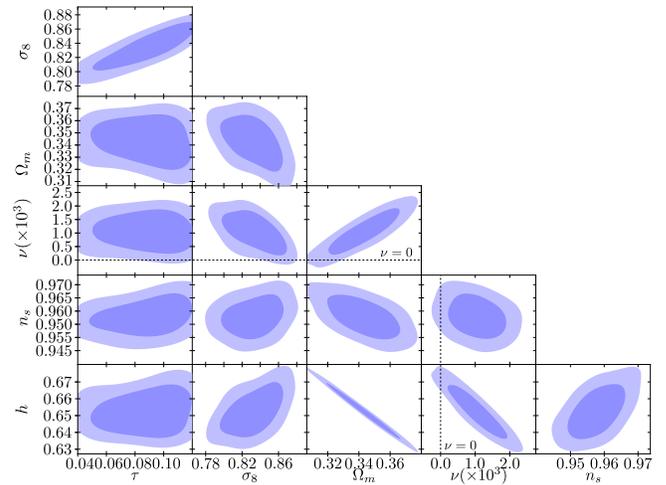}} \caption{2D likelihood contours for the pairs of sampled parameters for the $\Lambda(H)$CDM$_3$ model under the scope of the CMB data. The red line corresponds to $\nu=0$.}
	\label{fig:Q3}
\end{figure}

\begin{table*}
	\caption[]{The analysis results for models $\Lambda(H)$CDM$_1$, $\Lambda(H)$CDM$_2$ and $\Lambda(H)$CDM$_3$ using the data from Planck (TE,TT,EE+lowP), quoting the 68\% CL for the parameters.}
	\tabcolsep 7.5pt
	\vspace{1mm}
	\begin{tabular}{lrrrr} \hline \hline
		Model&\lcdm&$\Lambda(H)$CDM$_1$&$\Lambda(H)$CDM$_2$&$\Lambda(H)$CDM$_3$\\ \hline
		$\Omega_m$
		&$0.306\pm 0.018$
		&$0.309\pm 0.03$
		&$0.330\pm 0.03$
		&$0.342\pm 0.026$ \\
		$\sigma_8$ 
		&$0.83\pm 0.03$ 
		&$0.834\pm 0.04$
		&$0.822\pm 0.03$
		&$0.832\pm 0.035$\\
		$h$ 
		& $0.679\pm 0.012$ 
		& $0.678\pm 0.02$ 
		& $0.662\pm 0.02$ 
		& $0.653\pm 0.018$\\
		$\nu(\times 10^3)$ 
		&-
		& $0.04^{+0.71}_{-0.60}$ 
		& $0.59^{+1.0}_{-1.0}$ 
		& $1.0^{+0.8}_{-0.9}$ \\
		$n_s$
		&$0.964\pm 0.008$
		&$0.964 \pm 0.008$
		&$0.959\pm 0.008$
		&$0.958\pm 0.009$\\
		$\tau$
		&$0.068^{+0.037}_{-0.026}$ 
		&$0.071^{+0.038}_{-0.028}$
		&$0.064^{+0.028}_{-0.021}$
		&$0.086^{+0.030}_{-0.039}$\\
		\vspace{0.01cm}\\           
		\hline\hline
		\label{tab:cmb}
	\end{tabular}
\end{table*}

In Table I we show an overall presentation
of the current observational constraints imposed by the CMB 
spectrum, while in Figs. 1, 2 and 3 we plot the 
$1\sigma$ and $2\sigma$ 
contours in various planes for the explored $\Lambda(H)$CDM 
models. 

In particular, we find:

\begin{itemize}
	\item For the $\Lambda(H)$CDM$_{1}$ model we find 
	$\Omega_{m}=0.309\pm 0.03$, 
	$\sigma_{8}=0.834\pm 0.04$, 
	$h=0.678\pm 0.02$, 
	$\nu\times 10^3=0.04^{+0.71}_{-0.60}$, 
	$n_{s}=0.964 \pm 0.008$ 
	and $\tau=0.071^{+0.038}_{-0.028}$.
	
	\item In the case of $\Lambda(H)$CDM$_{2}$ model we obtain
	$\Omega_{m}=0.330\pm 0.03$, 
	$\sigma_{8}=0.822\pm 0.03$, 
	$h=0.662\pm 0.02$, 
	$\nu\times 10^3=0.59^{+1.0}_{-1.0}$, 
	$n_{s}=0.959\pm 0.008$ 
	and $\tau=0.064^{+0.028}_{-0.021}$.

	\item For the $\Lambda(H)$CDM$_{3}$ model:
	$\Omega_{m}=0.342\pm 0.026$, 
	$\sigma_{8}=0.832\pm 0.035$, 
	$h=0.653\pm 0.018$, $\nu\times10^3=1.0^{+0.8}_{-0.9}$, $n_{s}=0.958\pm 0.009$ 
	and $\tau=0.086^{+0.030}_{-0.039}$.

	\item In order to check the differences of the 
	$\Lambda(H)$CDM$_{i}$ models
	from the concordance $\Lambda$CDM case for which $\nu=0$. 
	Specifically, we find 
	$\Omega_{m}=0.306\pm 0.018$, 
	$\sigma_{8}=0.83\pm 0.03$, 
	$h=0.679\pm 0.012$, 
	$n_{s}=0.964\pm 0.008$ 
	and $\tau=0.068^{+0.037}_{-0.026}$.
	
	where we have quoted $1\sigma$ error bars.
	
\end{itemize}
We observe that the results from CAMB suggest that the best fit 
parameters of the dynamical vacuum models are in agreement with those of
$\Lambda$CDM case.
Moreover, 
the current observational constraints are compatible with those of
TT, TE, EE + lowE 
Planck 2018 data provided by Aghanim et. al. (2018) 
(see also Ade et. al. 2016), namely 
$\Omega_{m}=0.3166\pm 0.0084$, $\sigma_{8}=0.812\pm 0.0073$, 
$h=0.6727\pm 0.006$, $n_{s}=0.9649\pm 0.0044$ 
and $\tau=0.0544^{+0.0070}_{-0.0081}$.
Concerning the Hubble constant problem, namely 
the observed Hubble constant
$H_{0}=73.48 \pm 1.66$ Km/s/Mpc found by Riess et. al. (2018),  
is in $\sim 3.7\sigma$ tension with
that of Planck
$H_{0} = 67.36 \pm 0.54$ Km/s/Mpc (Aghanim et. al. 2018), we find that the
$H_{0}$ values extracted from the $\Lambda(H)$ models
are closer to the latter case (see also Sola et. al. 2017b).
Moreover our $H_{0}$ results are in agreement with those 
of Shanks, Hogarth \& Metcalfe (2018),  who found 
$H_{0}=67.6 \pm 1.52$ Km/s/Mpc using the GAIA parallax distances 
of Milky Way Cepheids.
However, the reader has to keep in mind that other 
studies support larger values of $H_{0}$ (cf. Lima \& Cunha2014; Beaton et al. 2016; Freedman 2017).

Moreover, we would also like to compare our results
with those of Yang et al. (2018), who have 
tested three scalar field DE models
against various datasets.
Particularly, for the Planck data 
the constraints of  Yang et al. (2018), [see Tables 1,2,4]  
stand as follows.
For the potential $V(\phi) \propto {\rm cosh}(\beta \phi)$ they found 
$\Omega_{m}=0.312\pm 0.009$, $\sigma_{8}=0.83\pm 0.014$, 
$h=0.6748\pm 0.006$, $\beta=4.32^{+1.5}_{-4.32}$, $n_{s}=0.9661\pm 0.0045$ 
and $\tau=0.081\pm 0.017$. 
In the case of 
$V(\phi) \propto 1+{\rm sech}(\alpha \phi)$ they obtained 
$\Omega_{m}=0.313\pm 0.009$, $\sigma_{8}=0.829\pm 0.013$, 
$h=0.6746\pm 0.0065$, $\alpha=4.77^{+5.23}_{-4.77}$, $n_{s}=0.966\pm 0.0046$ 
and $\tau=0.079\pm 0.017$. Lastly, for the potential 
$V(\phi)\propto [1+\delta (\phi/M_P)]^{2}$, where  
$M_P$ is the Planck mass, Yang et al. (2018), found
$\Omega_{m}=0.312\pm 0.0009$, $\sigma_{8}=0.829\pm 0.013$, 
$h=0.6747\pm 0.0064$, $\delta=4.843^{+5.157}_{-4.843}$, $n_{s}=0.966\pm 0.0044$ 
and $\tau=0.079\pm 0.017$. 
Obviously the constraints of the $\Lambda(H)$ models (see Table I) 
are compatible within $1\sigma$ with those of scalar field DE models.

\subsection{Combining with other probes} 

In this section we implement a joint likelihood analysis using 
SNIa from JLA sample 
(Betoule et. al. 2014), Baryonic Acoustic Oscillations (BAOs) 
(Blake et. al. 2011; Alam et. al. 2016) and measurements of 
$H_{0}$ Riess et. al. 2018, in order to place 
tight constraints 
on the corresponding parameter space of the models. The best fit parameters
are listed in Table II. 
In order to visualize the solution space 
in Figs.4, 5 and 6 we show the 1$\sigma$ and 2$\sigma$ 
contours for various planes. 
Specifically, the joint likelihood function peaks at (with the corresponding 68\% errors)

\begin{itemize}
	\item For the $\Lambda(H)$CDM$_{1}$ model:
	$\Omega_{m}=0.325\pm 0.008$, 
	$\sigma_{8}=0.808\pm 0.03$, 
	$h=0.661\pm 0.008$, 
	$\nu\times10^3=1.2^{+0.6}_{-0.5}$, 
	$n_s=0.959\pm 0.006$ 
	and $\tau=0.103^{+0.029}_{-0.033}$.
	
	\item For the $\Lambda(H)$CDM$_{2}$ model:
	$\Omega_m=0.291\pm 0.009$, 
	$\sigma_8=0.815\pm 0.02$, 
	$h=0.691\pm 0.008$, 
	$\nu\times10^3=-0.08^{+0.72}_{-0.78}$, 
	$n_{s}=0.967\pm 0.008$, 
	and $\tau=0.059^{+0.006}_{-0.008}$.
	
	\item For the $\Lambda(H)$CDM$_{3}$ model:
	$\Omega_{m}=0.295\pm 0.013$, 
	$\sigma_{8}=0.797\pm 0.04$, 
	$h=0.684\pm 0.01$, 
	$\nu\times 10^3=0.85^{+1.2}_{-0.9}$, 
	$n_{s}=0.962\pm 0.008$, 
	and $\tau=0.073^{+0.019}_{-0.024}$.
	
	\item Lastly, for the usual $\Lambda$CDM we find 
	$\Omega_{m}=0.298\pm 0.014$, 
	$\sigma_{8}=0.81\pm 0.01$, 
	$h=0.686\pm 0.010$, 
	$n_{s}=0.967\pm 0.008$ 
	and $\tau = 0.047^{+0.011}_{-0.006}$.
	
\end{itemize}

We find that the incorporation of more data sets 
via joint analyses improves
the fitting for all models, hence the current running vacuum models
are very efficient and in very good agreement with observations. 
Among the three $\Lambda(H)$ models the $\Lambda(H)$CDM$_{1}$ 
is the one with a small but non-zero deviation from the concordance
$\Lambda$CDM cosmology. Indeed we find that the deviation parameter $\nu$ is 
different from zero at $\sim 2.5\sigma$ level. 

First we verified that the combined 
constraints of the $\Lambda(H)$ models
(see Table II) are in agreement within $1\sigma$ errors with those 
of Yang et al. (2018), who considered the case 
of scalar field DE (see also Park \& Ratra 2018),
where these authors found: 
$\Omega_m=0.306\pm0.006$, $\sigma_8=0.83\pm0.013$, $h=0.6794\pm 0.0046$, $\beta=4.522^{+1.677}_{-4.522}$,
$n_s=0.9688\pm0.0038$ and $\tau = 0.085\pm 0.016$ 
[model $V(\phi) \propto {\rm cosh}(\beta \phi)$], 
$\Omega_m=0.306\pm0.006$, $\sigma_8=0.83\pm0.014$, $h=0.6797\pm 0.0049$, $\alpha=3.845^{+1.361}_{-3.845}$, $n_s=0.9691\pm0.0038$ and 
$\tau = 0.086\pm 0.018$ [model $V(\phi) \propto 1+{\rm sech}(\alpha \phi)$]
and finally
$\Omega_m=0.306\pm0.006$, $\sigma_8=0.83\pm0.013$, $h=0.6794\pm 0.0046$, $\delta=4.764^{+5.236}_{-4.764}$, $n_s=0.9688\pm0.0039$ and $\tau = 0.085\pm 0.017$ 
for the potential $V(\phi)\propto [1+\delta (\phi/M_P)]^{2}$.

Concerning the $\Lambda(H)$CDM$_{1}$ model our results can be compared
with those of Wang (2018), who combined SNIa/H(z)/BAO/CMB 
data. 
Notice that regarding the cosmological parameters $\{\Omega_{m},\nu,h\}$
our CMB$_{\rm shift}$/SNIa/BAO/$H_{0}$ constraints are similar (within $1\sigma$) 
with those of Sola et al. (2018)
(see also Gomez et. al. 2015, Sola et. al. 2017a, 2017b, and references therein) 
who found, combing 
cosmic chronometer, SNIa (JLA), CMB shift parameters and BAO data,
$\{\Omega_{m},\nu,h\}$=$\{0.304\pm 0.005,0.00014\pm 0.00103,0.684 \pm 0.007\}$,
$\{\Omega_{m},\nu,h\}$=$\{0.304\pm 0.005,0.00019\pm 0.00126,0.685 \pm 0.007\}$,
and $\{\Omega_{m},\nu,h\}$=$\{0.304\pm 0.005,0.0009\pm 0.0033,0.686 \pm 0.004\}$ for the
$\Lambda(H)$CDM$_{1}$, $\Lambda(H)$CDM$_{2}$ and $\Lambda(H)$CDM$_{3}$ 
models, respectively.
Regarding the $\Lambda(H)$CDM$_{2,3}$ models our constraints are in agreement 
with those of von Marttens et al. (2018) 
who used CMB (via CLASS code) in a joint analysis with 
$H_{0}$, SNIa, cosmic chronometer and BAO data.

Although our observational constraints are 
in qualitative agreement with previous studies (cf. Gomez et. al. 2015, Sola et. al. 2017a, 2017b, 2018)
we would like to spell out clearly the main reason why 
the results of the present work are novel. 
Indeed, to our knowledge, our analysis 
includes the Planck CMB power spectrum (via CAMB) in a large family of 
dynamical vacuum models and thus we can trace the 
Hubble expansion of the $\Lambda(H)$ models in the recombination 
era.

\begin{figure}
	\mbox{\epsfxsize=8.5cm \epsffile{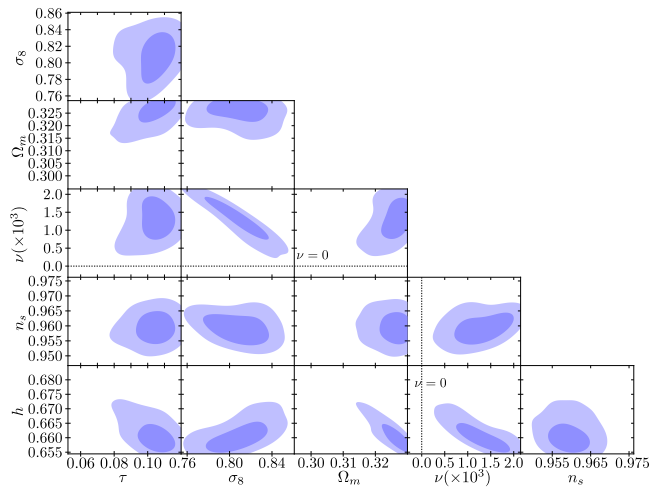}} \caption{Same as \ref{fig:Q1} but with the CMB/BAO/$H_0$/SNIa dataset.}
	\label{fig:Q1tot}
\end{figure}

\begin{figure}
	\mbox{\epsfxsize=8.5cm \epsffile{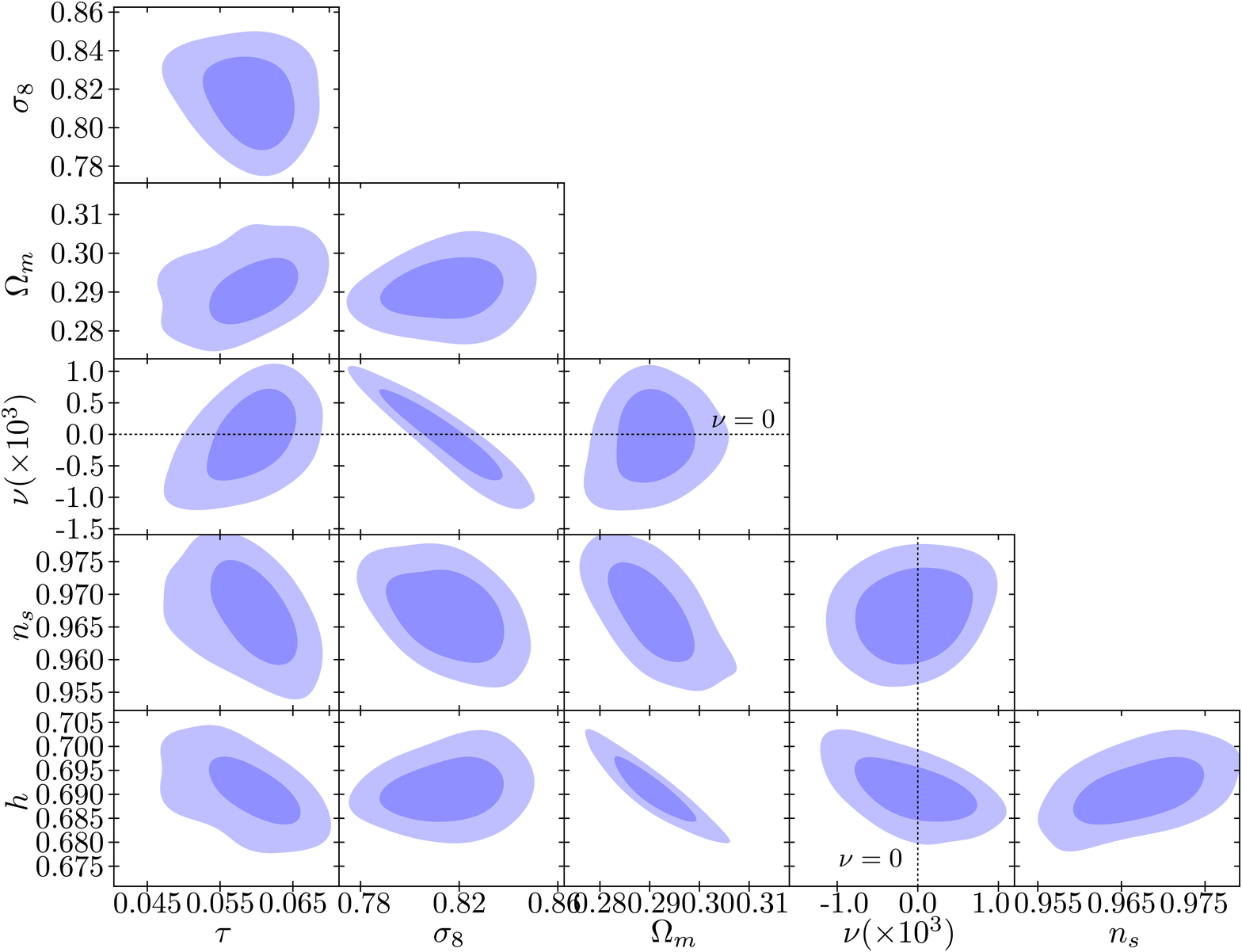}} \caption{Same as \ref{fig:Q2} but with the CMB/BAO/$H_0$/SNIa dataset.}
	\label{fig:Q2tot}
\end{figure}

\begin{figure}
	\mbox{\epsfxsize=8.5cm \epsffile{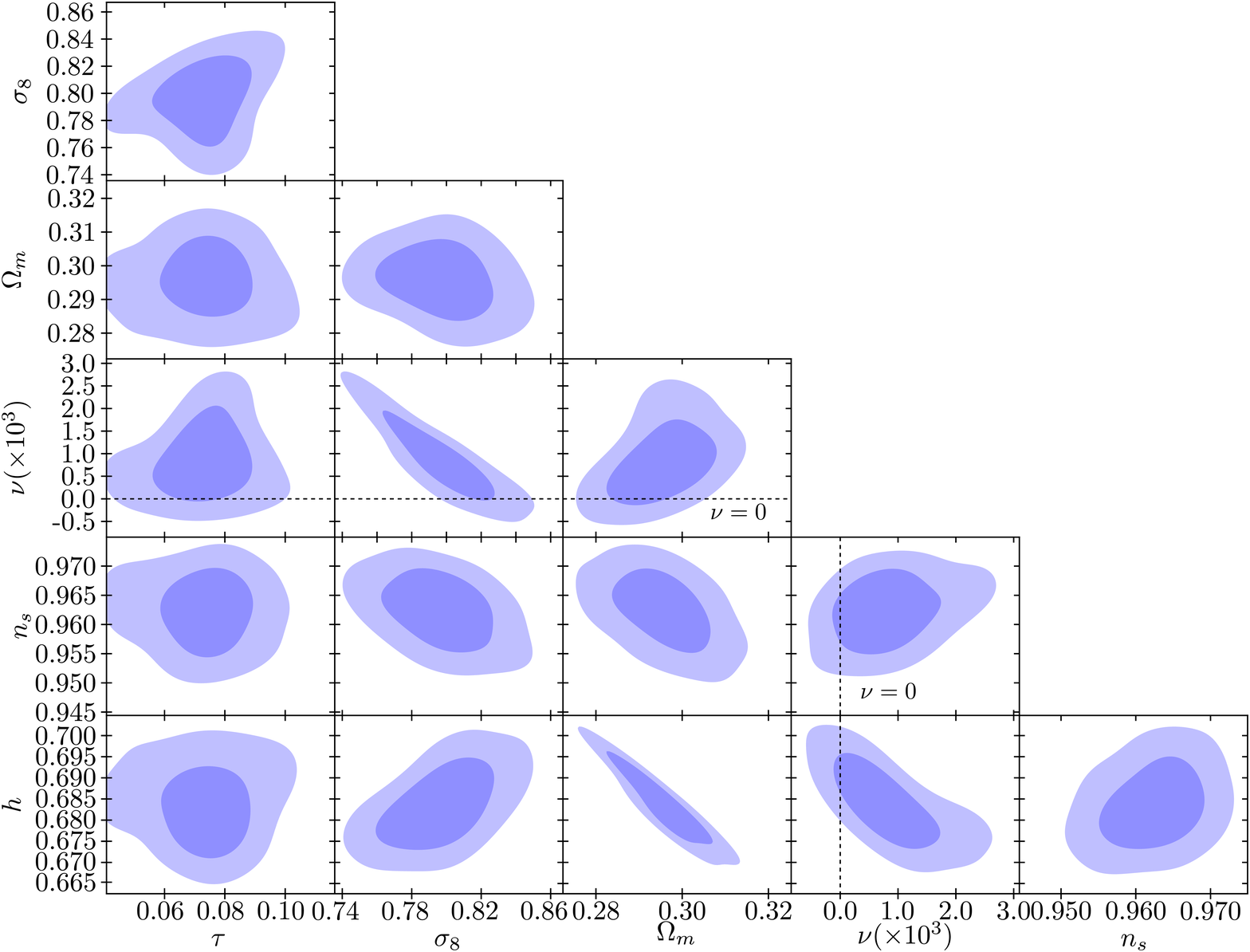}} \caption{Same as \ref{fig:Q3} but with the CMB/BAO/$H_0$/SNIa dataset.}
	\label{fig:Q3tot}
\end{figure}

\begin{table*}
	\caption[]{The analysis results for models $\Lambda(H)$CDM$_1$, $\Lambda(H)$CDM$_2$ and $\Lambda(H)$CDM$_3$ using a joint analysis of CMB data from Planck, BAO's, SNIa and $H_0$. In this table we quote the 68\% limits for the parameters.}
	\tabcolsep 7.5pt
	\vspace{1mm}
	\begin{tabular}{lrrrr} \hline \hline
		Model&\lcdm&$\Lambda(H)$CDM$_1$&$\Lambda(H)$CDM$_2$&$\Lambda(H)$CDM$_3$\\ \hline
		$\Omega_m$
		&$0.298\pm 0.014$
		&$0.325\pm 0.008$ 
		&$0.291\pm 0.009$ 
		&$0.295\pm 0.013$\\
		$\sigma_8$ 
		&$0.81\pm 0.01$
		&$0.808\pm 0.03$ 
		&$0.815\pm 0.02$ 
		&$0.797\pm 0.04$\\
		$h_0$ 
		&$0.686\pm 0.010$
		&$0.661\pm 0.008$ 
		&$0.691\pm 0.008$ 
		&$0.684\pm 0.010$ \\
		$\nu\times10^3$ 
		&-
		&$1.2^{+0.6}_{-0.5}$ 
		&$-0.08^{+0.72}_{-0.78}$ &$0.85^{+1.2}_{-0.9}$\\
		$n_s$
		&$0.967\pm 0.008$
		&$0.959\pm 0.006$
		&$0.967\pm 0.008$ 
		&$0.962\pm 0.008$\\
		$\tau$
		&$0.047^{+0.011}_{-0.006}$
		&$0.103^{+0.029}_{-0.033}$
		&$0.059^{+0.006}_{-0.008}$ 
		&$0.073^{+0.019}_{-0.024}$		\vspace{0.01cm}\\           
		\hline\hline
		\label{tab:joint}
	\end{tabular}
\end{table*}

\section{Conclusions}
\label{conclusions}
We extracted observational constraints on various 
dynamical vacuum models, using 
the CMB power spectrum from Planck. 
We used the most popular $\Lambda(H)$ models and in 
all of them we studied their deviation from the usual $\Lambda$CDM model
through a sole parameter. Modifying CAMB we found that the best fit 
parameters of the explored running vacuum models are 
in agreement with those of $\Lambda$CDM.
For completeness we combined the CMB spectrum in a joint analysis with other 
cosmological probes (SNe type Ia, BAOs, $H_{0}$) 
in order to place tight 
constraints on the cosmological parameters of the dynamical
vacuum models. We find that 
$\Lambda(H)$CDM$_{2}$ and $\Lambda(H)$CDM$_{3}$ do not 
show deviations from the $\Lambda$CDM case.
However, for the $\Lambda(H)$CDM$_{1}$ 
vacuum model,
we found a small but non-zero deviation 
from $\Lambda$CDM, where the confidence level is close
to $\sim 2.5\sigma$. This is an indication 
that dark energy could be dynamical.

\vspace{2cm}
\section*{Acknowledgements}
Spyros Basilakos would like to acknowledge 
support by the Research Center for Astronomy of the Academy
of Athens in the context of the program ``{\it Tracing the Cosmic
Acceleration}''. \\

\end{document}